\documentclass[prl,preprint,showpacs,floats,citeautoscript,amsmath,amssymb,floats]{revtex4}
\usepackage{graphicx}
\usepackage{dcolumn}
\usepackage{bm}

\def\be{\begin{equation}}
\def\ee{\end{equation}}
\def\bea{\begin{eqnarray}}
\def\eea{\end{eqnarray}}

\begin{document}
\title{\boldmath Screening of point charge impurities in highly anisotropic metals: application to $\mu^+$ spin relaxation in underdoped cuprates}
\author{Arcadi Shekhter, Lei Shu, Vivek Aji, D. E. MacLaughlin, C. M. Varma}
\affiliation{Department of Physics and Astronomy, University of California, Riverside, CA 92521}

\date{\today}

\begin{abstract}
We calculate the screening charge density distribution due to a point charge, such as that of a positive muon ($\mu^+$), placed between the planes of a highly anisotropic layered metal. In
underdoped hole cuprates the screening charge converts the charge density in the metallic-plane unit cells in the vicinity of the $\mu^+$ to nearly its value in the insulating state. The
current-loop ordered state observed by polarized neutron diffraction then vanishes in such cells, and also in nearby cells over a distance of order the intrinsic correlation length of the
loop-ordered state. This in turn strongly suppresses the loop-current field at the $\mu^+$ site. We estimate this suppressed field  in underdoped YBa$_2$Cu$_3$O$_{6+x}$ and
La$_{2-x}$Sr$_x$CuO$_4$, and find consistency with the observed 0.2--0.3~G field in the former case and the observed upper bound of $\sim$0.2~G in the latter case. This resolves the controversy
between the neutron diffraction and $\mu$SR experiments. The screening calculation also has relevance for the effect of other charge impurities in the cuprates, such as the dopants themselves.
\end{abstract}

\maketitle

In metals composed of metallic planes with negligible hopping from one plane to the next, the screening length for a point charge placed between the planes depends on the distance~$c$ between
the planes, the distance~$d$ of the charge to the planes, and the electronic compressibility of the planes. Through an elementary calculation based on the Thomas-Fermi method, we obtain the
screening charge distribution, and discuss how it is modified in strongly correlated metals.

\begin{figure}[ht]
 \begin{center}
 \includegraphics[width=0.45\textwidth]{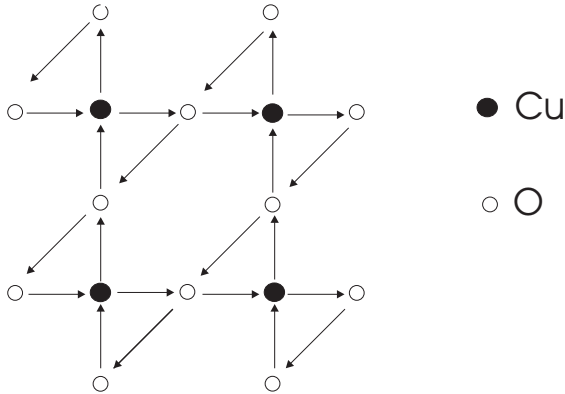}
 \caption{The orbital current pattern in the CuO$_{2}$ plane.}
 \label{fig:loop}
 \end{center}
\end{figure}

The immediate occasion for doing such a calculation is that sensitive muon spin relaxation ($\mu$SR) experiments in underdoped YBa$_2$Cu$_3$O$_{6+x}$ \cite{SonierScience,SonierPRB,sonier-pc} and
La$_{2-x}$Sr$_x$CuO$_4$ \cite{luke} detected at most only a small fraction of the magnetic field expected at the muon site given the magnetic order observed by polarized neutron scattering
\cite{FAQ,mook,greven}, and in the case of La$_{2-x}$Sr$_x$CuO$_4$ gave a null result \cite{luke}. Such a field is expected from proposed loop currents \cite{CMV1,CMV2} as illustrated in
Fig.~\ref{fig:loop}.

The discrepancy is huge: the expected field at the muon site is of order 100~G, while the $\mu$SR experiments yielded 0.2--0.3~G in YBa$_2$Cu$_3$O$_{6+x}$ \cite{sonier-pc} and no observable
field $\gtrsim 0.2$~G in La$_{2-x}$Sr$_x$CuO$_4$ \cite{luke}.

Loop-current magnetic order has been observed by polarized neutron diffraction experiments; the initial results on underdoped YBa$_2$Cu$_3$O$_{6+x}$ \cite{FAQ} have recently been verified in an
independent experiment \cite{mook}, and magnetic order of the same symmetry is also observed by neutron diffraction in underdoped HgBa$_2$CuO$_4$ \cite{greven}. In both of these compounds the
observed order is long-range, limited only by the resolution of the experiment. In underdoped La$_{2-x}$Sr$_x$CuO$_4$ only short-range order with a correlation length of about 10~\AA\ is
observed (around the same Bragg spots and with similar integrated intensity as the other compounds) \cite{bourges-pc}, due presumably to the larger charge disorder in this family of cuprates and
the resultant effects discussed in this paper \cite{LSCO}. The magnitude of the ordered moment is large, estimated to be 0.05--0.1\,$\mu_B$ per O--Cu--O plaquette at about 10\% doping,
decreasing to zero at the quantum critical point; the sublattice magnetization per unit cell is twice this value. Earlier dichroic ARPES results \cite{AK} on underdoped
Bi$_s$Sr$_2$CaCu$_2$O$_{8+\delta}$ samples revealed exactly the same symmetry of magnetic order. These experimental results were considered controversial \cite{BOR,campuzano, CMV3}, but the
neutron diffraction experiments lend them further credence. In underdoped YBa$_2$Cu$_3$O$_{6+x}$, a small dichroic signal characteristic of a time-reversal breaking order parameter has also been
observed \cite{kapitulnik} and may be associated with a small ferromagnetic order accompanying the main loop-current order \cite{aji3}.

On the basis of all these experiments, there are grounds now to believe that the loop-current order is universal in underdoped cuprates. It is therefore important to understand why $\mu$SR
experiments see little or no sign of the accompanying magnetic order.

The loop-current order can occur only in a limited range of electron density, from about 5\% to about 15\% deviation of charge density from the half-filled band; below $\sim$5\% doping the
materials are antiferromagnetic insulators, while the quantum critical point at which the loop current order ends is at $\sim$15\% doping. The experiments are done with positively charged muons
($\mu^+$), which unlike polarized neutrons are not innocent observers of a magnetic field. Their charge must be screened in the metal by repelling holes from their vicinity with an integrated
density equal to that of the $\mu^+$ charge. We show that this makes cells in the metallic planes in the vicinity of the muon acquire a local charge density characteristic of the insulator. This
of course removes the loop-current order locally. In addition to this effect, the magnitude of the loop-current order parameter is reduced around such cells over a distance of the order of the
intrinsic correlation length of the ordered state (several lattice constants). We calculate the field expected at the $\mu^+$ site when these effects are included, and find that it can be of the
order of the observed values in YBa$_2$Cu$_3$O$_{6+x}$ \cite{SonierScience,SonierPRB} and below the observed upper limit in La$_{2-x}$Sr$_x$CuO$_4$ \cite{luke}. Our elementary calculation and
conclusion also has applications to other properties, such as local one-particle spectra measured by STM and the effect of charged defects on transport properties in layered metals and
superconductors. We note that such screening does not occur in an insulator, where there is a gap to low energy charge excitations. Thus antiferromagnetism in the insulating state of the
cuprates is perfectly well observed in $\mu$SR experiments.

\begin{figure}[ht]
 \begin{center}
 \includegraphics[width=0.45\textwidth]{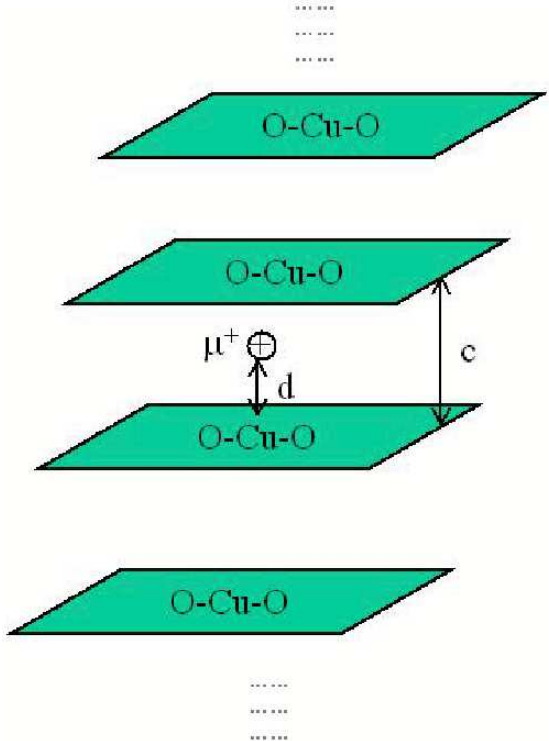}
 \caption{Schematic of CuO$_{2}$ planes and the $\mu^+$ site.}
 \label{fig:planes}
 \end{center}
\end{figure}

We use the model of metallic planes specified by the charge density of a typical underdoped cuprate, with each plane having a compressibility $dn/d\mu$ and a separation~$c$ between planes as
shown in Fig.~\ref{fig:planes}.

The positive charge is placed at a distance $d < c$ above one of the planes. We use the Thomas-Fermi method to calculate the potentials and the charge density changes. The potential $\phi(r,z)$
approximately follows the Poisson equation \bea -\nabla^2\phi({\bf r},z) = 4\pi \frac{e}{\epsilon} \big(\rho_{\rm ext}({\bf r},z) +\rho_{\rm ind}({\bf r},z)\big), \eea where $\epsilon$ has been
introduced to account for static polarizability in between the planes. The long wavelength limit of $\epsilon$ is known to be about $4$. The external charge is \be \rho_{\rm ext}({\bf r},z) =
\delta(z-d) \delta({\bf r}), \ee \noindent and the induced charge is given in the Thomas-Fermi approximation by \bea \rho_{\rm ind}({\bf r},z) = - \frac{e^{2}}{\epsilon}(dn/d\mu) \phi({\bf
r},z). \eea It is straightforward to show that the solution of these equations is such that the charge density $\rho_n(r)$ on the $n$th plane can be obtained by numerically evaluating
\begin{eqnarray}\label{eqn:chgind}
\rho_n(r) &=& \int \frac{\xi d\xi}{2\pi} J_0(\xi r) \int_{-\pi/c}^{\pi/c} \frac{dk}{2\pi} e^{ikz_n} \rho(k,\xi),\nonumber \\
\noalign{\noindent where the radial distance~$r$ is measured from directly below the muon,}
 \rho(k,\xi) &=& \frac{2\kappa_{\rm 2D}\,\sigma(k,\xi)}{ 1+ \displaystyle \frac{\kappa_{\rm 2D}}{\xi} \frac{\sinh{\xi c} }{ \cosh{\xi c} -\cos{kc}}}, \nonumber \\
 \noalign{\vskip5pt}
 \sigma(k,\xi) &=& \frac{1}{2\xi}
\frac{ e^{-ikc}\sinh \xi d + \sinh \xi(c-d) } { \cosh\xi c -\cos kc},\nonumber \\
\noalign{\vskip3pt \noindent and the response $\kappa_{\rm 2D}$ is defined by}
\kappa_{\rm 2D} &\equiv& 2\pi (e^2/\epsilon)(dn/d\mu).
\end{eqnarray}

In the general case, the screening charge distribution $\rho_n(r)$ can only be obtained by numerical evaluation. Some features of interest may however be noted immediately. For a charged
impurity near an isolated two dimensional conducting plane, the screening charge density decays as $1/r^3$ \cite{stern}. For large enough $ \kappa_{\rm 2D}$ (of the order or larger than the
effective mass enhancement divided by $\epsilon$ of about 1), the screening charge density is independent of $\kappa_{\rm 2D}$ and is determined solely by $c$ and $d$. For the usual situation
where $d$ is close to $c/2$, the screening is determined entirely by $c$. For $\kappa_{\rm 2D} \to 0$, the incompressible limit as in an insulator, there is no screening.

\begin{figure}[ht]
 \begin{center}
 \includegraphics[width=0.45\textwidth]{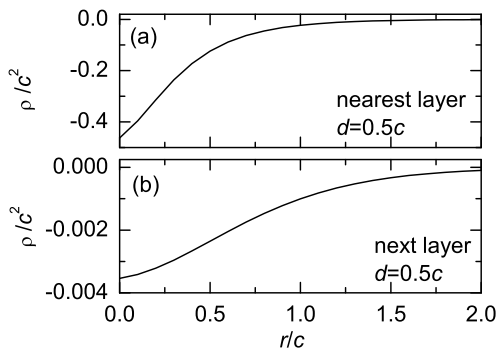}
 \caption{The screening charge density in the planes  in the Thomas-Fermi approximation for a muon centered between planes ($d = c/2$). (a): Nearest-neighbor plane. (b): Next-nearest-neighbor plane.}
 \label{fig:middle}
 \end{center}
\end{figure}

The calculated $\rho_n(r)$ in the planes in the immediate vicinity of the muon is shown in Fig.~\ref{fig:middle}.

More distant planes have negligible screening charge. A check on these calculations is that the integrated screening charge equals in magnitude the $\mu^+$ charge. We have taken $\kappa_{\rm 2D}
\approx 10/c$, estimated from an effective mass $\approx \epsilon \approx 4$. The results are insensitive to the value of $\kappa_{\rm 2D}$ beyond about $10/c$. For much smaller values of
$\kappa_{\rm 2D}$ than this, the screening distance goes up with a corresponding decrease of charge density immediately above and below the muon. This is as it should be, since for $\kappa_{\rm
2D} \to 0$ we should approach the insulating case.

There is an interesting general point about screening in strongly correlated material that is not directly included in the Fermi-Thomas approximation. In a parameter range as discussed below,
the Thomas-Fermi screening charge in the immediate vicinity of the charged defect is such that the charge density crosses over past the charge density in the insulator to effective local
electron doping. However, the compressibility of the insulator is zero, so that the screening charge in a cell cannot go beyond the insulating regime into the electron doping regime. Instead,
the charge density will locally stay close to zero doping, and the screening charge will be distributed more evenly outside the cell in the immediate vicinity as well as increased in more
distant planes. Within the above scheme this may be simulated by a {\it local} response $\kappa_{\rm 2D}[\rho(r)]$, which goes to zero when $\rho(r)$ is in the insulating regime. This has the
effect of flattening the charge distributions compared to those shown in Fig.~\ref{fig:middle} and therefore increasing the size of the screening charge.

We now apply these calculations to the parameters of underdoped La$_{2-x}$Sr$_x$CuO$_4$. In this compound the planes are separated by $c \approx 6.7$~\AA\ and the in-plane lattice constant is $a
\approx 3.8$~{\AA}\@. From Fig.~\ref{fig:middle}, for the case $d=0.5c$ ($\mu^+$ in the middle of the two planes), the integrated screening charge density in the cell immediately above and below
the $\mu^+$ changes by about 10\%, and in each of the neighboring 8 cells it changes by about 5\%. There is negligible screening charge farther away. At a typical underdoping concentration of,
say, 10\%, this places the cells directly underneath definitely in the insulating regime and their neighbors close enough that the proposed orbital order is strongly affected in them from this
effect alone. As noted above, incompressibility of the insulating regime increases the size of the screening region beyond this estimate. In addition, the suppressed loop-current order requires
a significant additional distance to recover as shown below. Thus even the conversion of one cell to the insulating composition is enough to make a large $\mu^+$ field unlikely.

\begin{figure}[ht]
 \begin{center}
 \includegraphics[width=0.45\textwidth]{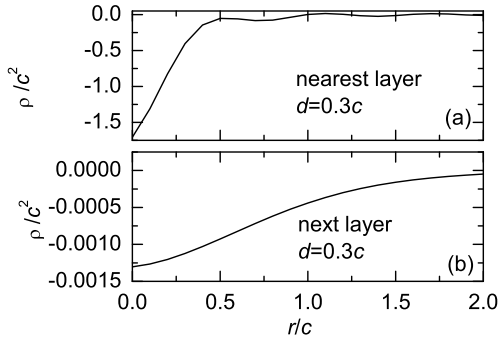}
 \caption{The screening charge density in nearby planes in the Thomas-Fermi approximation, with a large asymmetry in the placement of the muon ($d = 0.3c$). (a): Nearest neighbor plane. (b): Next nearest  neighbor plane.}
 \label{fig:asym}
 \end{center}
\end{figure}

For $d=0.3c$, which is a large asymmetry in the placement of the muon, the screening charge density in the nearest and the next nearest neighbor plane are shown in Fig.~\ref{fig:asym}.

In the Thomas-Fermi approximation the charge in the nearest cell in the closest plane changes by about 30\%, and the charge in in the 8 neighboring cells changes by about 3\%. The numbers in the
plane in the other side are reduced correspondingly. As discussed above, the screening charge distribution in the closest plane must be larger and more smoothly varying than in this
approximation.

Consider now the effect on the spatial variation of the loop-current order parameter $\Upsilon$ on reducing it to zero by the above effect in the metallic cells close to the muon. The amplitude
of the order parameter can only recover to its full value over the characteristic coherence length $\xi$ of the order parameter $\Upsilon(r)$ beyond the cells where $\Upsilon$ vanishes. The
spatial dependence of $\Upsilon$ is determined by \be \xi^2 \nabla^2\Upsilon ({\bf r}) = \frac{\partial F_0}{\partial \Upsilon(\bf r)}, \ee where $F_0$ is the free energy of the homogeneous
loop-order phase, with the boundary condition that $\Upsilon$ is zero in the lattice cells close to the muon. This equation also occurs in the context of the variation of the order parameter in
the core of a vortex except that here the boundary condition $\Upsilon({\bf r}{=}0)=0$ is replaced by $\Upsilon({\bf r})=0$ for ${\bf r} \lesssim a_0$, the radius in which the local charge
density is in the insulating range. A dimensional argument suggests that at low temperatures $\xi \approx a ({E_e}/{E_{c\ell}})^{1/2}$, where $a$ is the in-plane lattice constant, $E_e$ is a
typical electronic energy (e.g., the conduction electron bandwidth $\sim 1$~eV) and $E_{c\ell}$ is the condensation energy of the loop-order phase. Even if we take for the latter a value as high
as $0.1$ eV, $\xi$ is about $3a$. Near the ordering temperature $\xi(T)$ diverges.

Thus the loop-current order parameter is suppressed for a number of lattice constants beyond the cells where it is truly zero. Since one or more cells adjacent to the muon have zero order
parameter, in estimating the magnetic field at the $\mu^+$ site it is reasonable to exclude near-neighbor cells as well as the cell immediately below or above the muon. Then the estimated field
at the $\mu^+$ site is consistent with experiment as shown below.

The calculation is done by setting up an orthorhombic lattice, placing a $\mu^+$ site and local dipole moments in a magnetic structure in the unit cell, and summing over the local moments out to
a given radius. The muon is placed $1$~\AA\ below the apical oxygen as suggested by some experiments \cite{muonexperts}, although there seems to be no clear consensus on this location
\cite{SonierScience}. With no change in the loop-current order parameter due to screening of the $\mu^+$ charge, the magnetic field at the $\mu^+$ site for a magnetic order with a moment of
$0.05\, \mu_B$ per triangular plaquette placed at the centroid of the triangles is about 240~G for La$_{2-x}$Sr$_x$CuO$_4$ (a value of 41~G is obtained from a more precise calculation of the
field due to the loop current pattern \cite{luke}.) If we now exclude 9 cells in the immediate vicinity of the $\mu^+$ site in each of the two planes, i.e., $a_0+\xi =2$ lattice parameters
(7.6~\AA), the calculated $\mu^+$ magnetic field is about 3.5~G. When the next neighboring set of cells are excluded, i.e. $a_0+\xi =3$ lattice parameters (11.4~\AA) the field is about 1~G,
while for $a_0+\xi =4$ lattice parameters (15.2~\AA) it is about 0.45~G. Given the uncertainty of the muon site, moving it $15\%$ closer to the apical Oxygen leads to a field of 4.35~G, 1.6~G
and 0.69~G when screened to two, three and four lattice spacings respectively.  The above discussion does not take into account the small correlation length found for the loop order in
La$_{2-x}$Sr$_x$CuO$_4$ \cite{bourges-pc}, which could further reduce the estimated field.

For the case of YBa$_2$Cu$_3$O$_{6+\delta}$ we find an unscreened $\mu^+$ field of about 290~G for the muon at $1$~\AA\ below the apical oxygen and the same loop-current magnetic moment as
above. The loop-current reduction due to screening reduces this to $\sim$3~G when only the cells immediately above and below the muon are excluded, and to $\sim$0.5~G when nearest- and
next-nearest neighbor cells are also excluded.

The precise values obtained from these calculations vary with details that are not accurately known, such as the muon position. The principal point is that the $\mu^+$ field due to loop currents
is reduced by more than two orders of magnitude due to the inescapable effects of charge screening.

For metallic doping levels the lattice cells near a $\mu^+$ site with charge density close to that of the insulator will acquire local spin moments at Cu sites. These
anti\-ferro\-mag\-netically-coup\-led Cu spins will not freeze because of thermal and quantum fluctuations in such small regions, but the $\mu^+$-induced Cu-spin cluster should act as a
paramagnetic center and thus might be observable in $\mu$SR experiments. At temperatures well above the superconducting transition $\mu$SR measurements give some evidence for weak dynamic
relaxation in YBa$_2$Cu$_3$O$_{6+x}$ \cite{sonier2} and La$_{2-x}$Sr$_x$CuO$_4$ \cite{sonier2,Watanabe}; in La$_{2-x}$Sr$_x$CuO$_4$ a $\sim$2~G field fluctuating at a very slow rate (0.1~MHz)
was inferred \cite{Watanabe}. Although a Cu-spin cluster might be expected to fluctuate slowly, at present its dynamics are not sufficiently well understood to allow comparison with experiment.

Such a cluster should, however, have a paramagnetic response appropriate to its net moment, which should lead to a muon Knight shift proportional to $1/T$ when an external magnetic field is
applied. This temperature dependence is quite different from that expected from bulk susceptibility measurements. Accurate muon Knight shift experiments have not yet been carried out
\cite{sonier-pc}. Transverse-field $\mu$SR experiments have, however, detected inhomogeneous static distributions of $\mu^+$ fields with widths proportional to the applied field \cite{sonier2}.
E.g., muons see a field-induced distribution of inhomogeneous magnetic fields with a spread of about 2~G at a field of 3 T\@. Suppose there is a $1\ \mu_B$ moment produced in the insulating cell
above and below the muon. With a Curie susceptibility, it produces an extra magnetic field at the $\mu^+$ site of $O(\mu_BH/T)\mu_B/|r|^3$, where $|r|$ is a typical distance to the muon of about
5~\AA\ and $H$ is the applied field. At a field of 3~T and a temperature 100~K, this field is about 100~G\@. If we allow a distribution of the detailed position of the muons due to local
strains, a field distribution with a width $\propto H/T$ is expected in the normal state. Such behavior is consistent with the observations \cite{sonier2}. The observed width of about 2~G at
$H/T = 3 \times 10^{-3}$~T/K is also consistent with our ideas. To test them further, a $\mu^+$ Knight shift $\propto 1/T$ should be looked for.

\begin{acknowledgments}
We are grateful to P. Bourges, A. Kapitulnik, and J. Sonier for useful discussions. This work was supported in part by the U.S. NSF, Grant no.~0422674.
\end{acknowledgments}


\begin{thebibliography}{21}
\bibitem[{LSC()}]{LSCO}
\bibinfo{note}{One may argue that La$_{2-x}$Sr$_x$Cu$_4$ is therefore perhaps
  not the ideal system in which to look for local fields. But $\mu$SR
  experiments, being local probes, should see a large effect even in the
  presence of loop-current disorder.}
\end{thebibliography}


\end{document}